# Further evidence for a quasar-driven jet impacting its neighbour galaxy: the saga of HE0450−2958 continues

D. Cs. Molnár,[1][⋆] M. T. Sargent,[1] D. Elbaz,[2] P. P. Papadopoulos[3,4] and J. Silk[2,5,6,7]

[1]*Astronomy Centre, Department of Physcis and Astronomy, University of Sussex, Brighton BN1 9QH, UK*
[2]*Laboratoire AIM, CEA/DSM-Université Paris Diderot, IRFU/Service d'Astrophysique, Bât.709, CEA-Saclay, F-91191 Gif-sur-Yvette Cédex, France*
[3]*School of Physics and Astronomy, Cardiff University, Queen's Buildings, The Parade, Cardiff CF24 3AA, UK*
[4]*Department of Physics, Section of Astrophysics, Astronomy and Mechanics, Aristotle University of Thessaloniki, Thessaloniki 54124, Greece*
[5]*Institut d'Astrophysique de Paris (UMR7095: CNRS and UPMC-Sorbonne Universities), F-75014 Paris, France*
[6]*Department of Physics and Astronomy, The Johns Hopkins University Homewood Campus, Baltimore, MD 21218, USA*
[7]*BIPAC, Department of Physics, University of Oxford, Keble Road, Oxford OX1 3RH, UK*

## ABSTRACT

HE0450−2958, an interacting quasar–starburst galaxy pair at $z = 0.285$, is one of the best-known examples of strong star formation activity in the presence of a quasar-driven jet. We present new multiband Karl G. Jansky Very Large Array-imaging covering 1–6 GHz and reaching an angular resolution of up to 0.6 arcsec (a sixfold improvement over existing radio data). We confirm the previous detection of a spatially extended radio component around the quasar, indicating that there is ongoing star formation activity in the quasar host galaxy. For the first time, we directly detect a jet-like bipolar outflow from the quasar aligned with its companion star-forming galaxy (SFG) and several blobs of ionized gas in its vicinity identified in previous studies. Within the companion SFG, we find evidence for a flattening of the synchrotron spectral index towards the point of intersection with the jet axis, further suggesting that the outflow may actually be impacting its interstellar medium. We discuss two possible mechanisms that could have triggered the starburst in the companion SFG: a wet–dry merger with the quasar and jet-induced star formation. While triggering through interaction-driven gas dynamics cannot be excluded with current data, our new observations make HE0450−2958 a strong candidate for jet-induced star formation, and one of the rare links between local systems (like Minkowski's Object or Centaurus A) and the high-$z$ regime where radio-optical alignments suggest that this phenomenon could be more common.

## 1 INTRODUCTION

Observations show that the mass of the central supermassive black hole of local galaxies correlates with their bulge luminosity (Kormendy & Richstone 1995; Magorrian et al. 1998), bulge mass (Kormendy & Gebhardt 2001; McLure & Dunlop 2001, 2002; Marconi & Hunt 2003; Ferrarese et al. 2006) and velocity dispersion (Ferrarese & Merritt 2000; Gebhardt et al. 2000). How these scaling relations are established is not well understood. It could simply be the outcome of stochastic growth through merging in a hierarchical universe (e.g. Peng 2007; Jahnke & Maccio 2011), or the joint outcome of gas consumption by similar relative proportions for star formation (SF) and black hole growth (e.g. Mullaney et al. 2012). A further possibility is a causal connection, e.g. through feedback pro-cesses: current models describing galaxy formation and evolution tend to use various prescriptions for feedback from active galactic nuclei (AGNs) to match their predictions to observations (Silk & Mamon 2012 and references therein), especially for high-mass galaxies. Therefore, studying the effects of AGN activity on its surroundings is one of the keys to understanding galaxy formation and evolution.

### 1.1 Negative and positive feedback

Feedback due to AGN activity is often used to explain quenching of SF in galaxies via radiation and mechanical feedback interacting with the interstellar medium (ISM) of the host galaxy and the intracluster gas surrounding it. Two modes of negative AGN feedback are commonly discussed in the literature: a radiative mode ('quasar' or 'cold' mode) for very luminous, fast-accreting AGN generating a radiation pressure that is capable of expelling gas from the host galaxy, and a kinetic mode ('radio' or 'hot' mode) whereby slowly accreting AGN drive jets and cocoons that heat intracluster gas and inhibit cooling and accretion on to the host galaxy from the circumgalactic medium. For a more in-depth review of the subject, see Fabian 2012 and references therein. Negative feedback is invoked to explain the origin of red and dead galaxies in the local Universe. However, even if quasars ultimately stop SF in galaxies, they may also act as a trigger at an earlier stage of their evolution, through positive feedback due to radio jets and turbulent pressure (see e.g. Begelman & Cioffi 1989; Silk 2005, 2013; Silk & Norman 2009; Nesvadba et al. 2011). Elbaz et al. (2009) have proposed that quasar-driven jets can play an active role in galaxy formation through positive feedback. The compact nature of nuclear starbursts, simulated in Gaibler et al. (2012), is comparable to that now being observed with ALMA, e.g. in Oteo et al. (2016a) via extreme, Arp 220-like, SF densities. While there is no direct evidence for an AGN, and associated triggering of a circumnuclear disc (Oteo et al. 2016b), the case for an AGN being required in the analogous case of Arp 220 was presented by Tunard et al. (2015) from chemical evidence.

⋆ E-mail: d.molnar@sussex.ac.uk

A promising candidate for observing both positive and negative feedback in action is HE0450−2958.

### 1.2 HE0450−2958 – a peculiar object with a history

HE0450−2958 is an optically bright quasar ($M_V = -25.8$) originally classified as a Seyfert I galaxy due to its far-infrared (FIR) colours (de Grijp, Lub & Miley 1987) and its optical spectra (Merritt et al. 2006) at the redshift of $z = 0.285$. Optical images with *Hubble Space Telescope* (*HST*) revealed a double system with the quasar and a companion star-forming galaxy (companion SFG) with ∼7 kpc projected separation (Magain et al. 2005, M05 henceforth), which is located at the same redshift as the quasar (within the observational uncertainties of the spectroscopy presented, e.g. in Letawe, Magain & Courbin 2008). After deconvolving the *HST* image, M05 also found an emission line 'blob' at the redshift of the quasar with small projected separation. M05 proposed that it is an AGN-ionized gas cloud; however, more recently, Letawe & Magain (2010) found near-IR continuum emission towards the source, suggesting that it is an off-centre, bright and very compact host galaxy.

HE0450−2958 spurred significant interest when M05 estimated the black hole mass for the quasar and turned an upper limit for the host galaxy luminosity that was five times fainter than that expected based on the Magorrian relation between black hole mass and bulge luminosity (Magorrian et al. 1998). The finding triggered a debate about the existence of so-called naked quasars, i.e. quasars without a host galaxy. This hypothesis is no longer strongly favoured in the case of HE0450−2958 following (a) publication of a revised black hole mass estimate by Merritt et al. (2006), which is 10 times lower than the one in M05, and (b) evidence of extended radio continuum emission from SF activity in the quasar host galaxy (Feain et al. 2007). However, the quasar is still not placed firmly on the Magorrian relation and the quasar host galaxy properties remain poorly known.

The second main component of the HE0450−2958 system, the companion SFG, has attracted attention in its own right. It was initially thought to be a ring galaxy (Boyce et al. 1996), but later M05 discovered that it hosts a highly dust-obscured central region that produces the ring-like appearance at optical wavelengths. Based on low-resolution radio continuum imaging, Feain et al. (2007) proposed that the quasar and the companion SFG are bridged together by a radio jet (as seen in Fig. 1) that was suggested to induce the in-tense SF (∼340 M$_{\odot}$ yr$^{-1}$; Elbaz et al. 2009) in the companion SFG and turn it into an ultraluminous infrared galaxy (ULIRG; Elbaz et al. 2009). However, due to the poor angular resolution (∼3 arcsec or 13 kpc at the distance of HE0450−2958, see Table 1) of the Australia Telescope Compact Array (ATCA) imaging, they could provide only indirect evidence with the detection of the opposite lobe. The presumed interacting outflow component remained undetected, since it was blended with emission from the companion SFG. Following this study, Elbaz et al. (2009) suggested that HE0450−2958 may represent an early phase in a scenario of 'quasar-induced galaxy formation'. While it has been debated whether the active SF in the companion is due to a merger event between the two objects or actual jet-induced SF – as found in other systems such as Minkowski's Object (Croft et al. 2006) – high excitation optical lines have been found in the companion with hints of shock-induced origins, suggesting that its gas was undergoing collision (Letawe et al. 2008). Meanwhile, Papadopoulos et al. (2008) found molecular gas associated mainly with the companion SFG, not the quasar, suggesting that the AGN could provide negative feedback for its own host galaxy. This makes HE0450−2958 an example highlighting the full diversity of AGN feedback-related phenomena in one system (similar to e.g. Cresci et al. 2015).

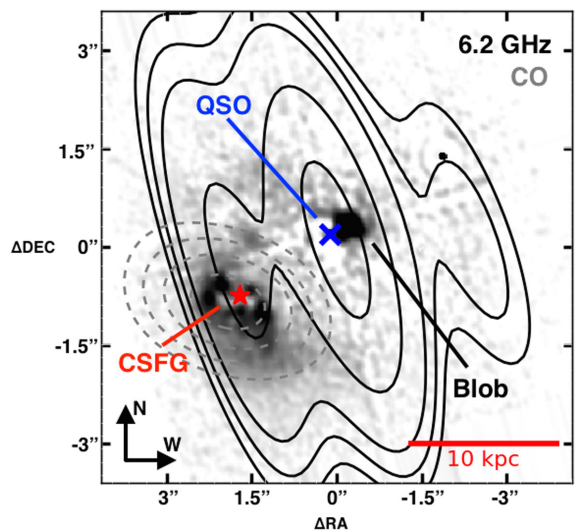

**Figure 1.** PSF-deconvolved *HST* optical image of HE0450−2958 from M05. The red star and the blue cross show the positions of the companion SFG (CSFG) and quasar (QSO), respectively. An emission line 'Blob' is visible directly to the west of the quasar. The black contours show the triple-component 6.2 GHz radio flux distribution measured with the ATCA (Feain et al. 2007; contour levels are $5\sigma$, $7\sigma$, $10\sigma$, $20\sigma$ and $35\sigma$). Grey dashed contours represent the CO $J = 1$–$0$ line detection obtained with ATCA by Papadopoulos et al. (2008, contour levels are $7\sigma$, $9\sigma$, $11\sigma$ and $13\sigma$). See text of Section 1.2 for further discussion.

In these papers, various scenarios were proposed to explain the observations, e.g. a merger of the massive companion SFG and a dwarf elliptical with an AGN (Papadopoulos et al. 2008), quasar-jet-induced SF in the companion SFG (Elbaz et al. 2009) or a highly asymmetric host with an offset active nucleus (Letawe & Magain 2010). Our new radio observations mainly aim to test the quasar-induced large-scale SF hypothesis.

Throughout this paper, we use a flat $\Lambda$CDM cosmology with $\Omega_M = 0.31$ and $H_0 = 67.77$ km Mpc$^{-1}$ s$^{-1}$ (Planck Collaboration XVI 2014). Star formation rates (SFRs) and stellar mass values reported assume a Salpeter initial mass function.

**Table 1.** Summary of JVLA observations and image properties (lines 1–3) of HE0450−2958 ATCA data obtained by Feain et al. (2007). Image noise was measured in emission-free regions near the phase centre. The resolution column provides the FWHM of the semimajor and semiminor axis of the restoring beam. Restoring beam position angles (PA) are measured counterclockwise from the positive y-axis.

| Observation | Frequency (GHz) | Resolution (arcsec × arcsec) | PA (deg) | $\sigma_{rms}$ ($\mu$Jy beam$^{-1}$) | On-source time (min) |
|---|---|---|---|---|---|
| JVLA BnA array | 1.5 GHz | 3.52 × 2.71 | −54 | 16 | 72 |
| JVLA A array | 1.8 GHz | 0.84 × 2.64 | 16 | 32 | 21 |
| JVLA A array | 5 GHz | 0.93 × 0.37 | −3 | 11 | 108 |
| ATCA | 6.2 GHz | 5.68 × 1.95 | 16.1 | 40 | – |
| ATCA | 8.6 GHz | 4.07 × 1.25 | 14.9 | 50 | – |

## 2 OBSERVATIONS AND DATA REDUCTION

### 2.1 Observations

We observed HE0450−2958 on 2011 September 3 and 2012 September 11 (project codes 10C-123 and 12B-192, respectively; PI: Sargent) with the Karl G. Jansky Very Large Array (JVLA) in the *L* and *C* bands using 26 antennas. Three setups were chosen to achieve distinct science goals.

(i) Simultaneous A-array observations in two *C*-band spectral windows centred at 4 and 6 GHz produced high-resolution images probing the small-scale structure of the GHz radio emission from the HE0450−2958 system. With an average angular resolution of 0.37 arcsec × 0.93 arcsec, these images allow us (a) to identify the eastern lobe of a potentially bipolar outflow from the quasar by spatially separating it from the emission from the companion SFG (see Section 3.2) and (b) to study variations of the radio spectral index in the companion SFG (see Section 3.3).

(ii) A-array imaging at 1.8 GHz (angular resolution 0.84 arcsec × 2.64 arcsec) can, for the first time, resolve the three main radio components of the HE0450−2958 system and determine their spectral properties using our higher frequency observations. When combined with the ATCA *X*-band data from Feain et al. (2007), our *L*- and *C*-band data sample the radio spectral index of HE0450−2958 in approximately every 2 GHz between 1.8 and 8.6 GHz (i.e. a total frequency baseline of 6.8 GHz).

(iii) Using the full 1-GHz frequency coverage of the *L*-band receivers in the more compact BnA configuration and a long integration time, we obtained lower resolution (3.52 arcsec × 2.71 arcsec) but high-sensitivity images to search for faint synchrotron emission from, e.g. ionized region produced by quasar jet wobbling or displaying SF activity triggered by the quasar jet (see discussion in Elbaz et al. 2009).

Integration times (see Table 1) were based on the total fluxes of the currently known radio components associated with HE0450−2958, as measured by Feain et al. (2007) with ATCA. Feain et al. (2007) achieved detections at an 8σ (*C* band, 6.2 GHz) and 6σ (*X* band, 8.6 GHz) significance level with ATCA. The targeted sensitivity and angular resolution of our observations were chosen to significantly improve the imaging quality by achieving S/N ∼ 10 in each resolution element on the faintest of the previously observed radio components associated with HE0450−2958, i.e. the western lobe (the opposite lobe of a putative bipolar outflow that has been suggested to have triggered SF in the companion SFG).

Our BnA-configuration *L*-band follow-up involved 72 min of on-source observations with 1 s integrations. 16 spectral windows with

64 channels each were used to cover the entire *L* band from 1 to 2 GHz. All data from antenna 10 had to be flagged during these wideband observations due to its *L*-band receiver being removed for repairs. The total available bandwidth during the *L*- and *C*-band observations in configuration A was limited due to the JVLA being in the VLA to EVLA transition phase at the time of observation. The *L*-band observations in configuration A used a total bandwidth of 256 MHz with 2-MHz frequency resolution centred at 1.8 GHz to avoid the regions most affected by strong radio frequency interference (RFI) and to maximize the angular resolution achievable in the *L* band. The total integration time of these observations was 21 min with 1 s integration intervals. For our *C*-band observations, the correlator was configured to deliver two spectral windows centred at 4036 and 5936 MHz, each providing 64 × 2 MHz channels for a total bandwidth of 256 MHz. In this setup, we observed the source for 108 min also with 1 s integration intervals.

During each scheduling block (SB), we followed the same observing strategy: alternating scans of our science target, HE0450−2958, and the phase calibrator J0453−2807 that also served as bandpass calibrator. For flux calibration, we used a single pointing on J0137+3309 (3C 48) at the end of each SB.

### 2.2 Calibration

Flagging of bad data and calibration were done using standard procedures as implemented in the VLA Calibration pipeline[1] in CASA.[2] Hanning smoothing was applied to all data to ensure a more effective detection and removal of RFI. For the more strongly affected *L*-band data, the reliability of the pipeline's automatic RFI flagging with the RFLAG task was tested in several spectral windows on phase/bandpass and flux calibrator data by comparison to manually flagged data. Thirty-eight per cent of the visibilities at 1.5 GHz, ∼22 per cent of those at 1.8 GHz and ∼17 per cent of the visibilities in the *C* band, were flagged. After calibration, all antenna-based amplitude and phase solutions were inspected and found to be satisfactory. The flux calibrator (3C 48) was imaged in each band and its flux measured to be within 6 per cent (1.5 GHz), 6 per cent (1.8 GHz), and 10 per cent (5 GHz) of the accepted values.

### 2.3 Imaging

Imaging was carried out in CASA using the CLEAN task in multiscale multifrequency synthesis mode (MS-MFS; Rau & Cornwell 2011)

---

[1] https://science.nrao.edu/facilities/vla/data-processing/pipeline
[2] Common Astronomy Software Applications: http://casa.nrao.edu

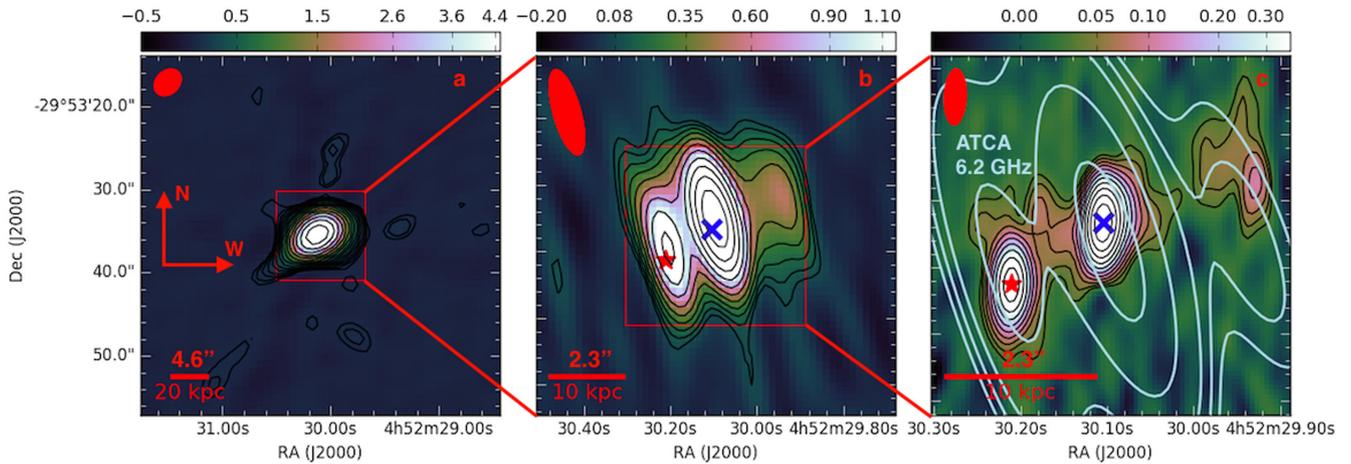

**Figure 2.** Overview of HE0450−2958 multifrequency data described in this paper. Left: JVLA 1.5 GHz (*L* band) wideband image obtained in BnA configuration. The total cut-out size is 43.3 arcsec × 43.2 arcsec (188 × 188 kpc²). Down to an rms noise level of 16 mJy beam⁻¹ (corresponding to a 3σ upper limit on the SFR of 36 M⊙ yr⁻¹), we find no evidence for synchrotron emission from extended emission line regions in the vicinity of HE0450−2958. The red square shows the region plotted in the central panel of the figure. Middle: JVLA 1.8-GHz image obtained with A-array configuration (cut-out size: 10.8 arcsec × 10.8 arcsec or 47 × 47 kpc²). The same triple structure as observed at higher frequencies is visible. The red square shows the region plotted in the right-hand panel. Right: JVLA 5-GHz (*C* band) image obtained in A-array configuration (cut-out size: 5.4 arcsec × 5.4 arcsec or 23.5 × 23.5 kpc²). Contours overlaid in light blue are from the lower resolution 6.2-GHz ATCA image in Feain et al. (2007). ATCA contour levels are 0.2 mJy beam⁻¹ (5σ), 0.28 mJy beam⁻¹ (7σ), 0.4 mJy beam⁻¹ (10σ), 0.8 mJy beam⁻¹ (20σ) and 1.4 mJy beam⁻¹ (35σ). Black contours for flux distributions measured with the JVLA start at 3σ, all other contours are $\sqrt{2}^n$ of it. At the highest resolution, the triple structure in the central panel is resolved into a more complex radio flux distribution. Black contours for flux distributions measured with the JVLA start at 3σ, all other contours are $\sqrt{2}^n$ of it in all three panels. Red ellipses in all panels denote the FWHM of the JVLA restoring beam at the respective frequencies. The red star and the blue cross show the *HST* optical image positions of the companion SFG and quasar, respectively (see Fig. 1).

and Briggs weighting with a robust parameter of 0.5. When combining the two spectral windows centred at 4 and 6 GHz in the *C* band, we were able to model the frequency dependence of visibilities (nterms = 2 in the MS-MFS mode), resulting in a spatially resolved spectral index map additional to the standard full Stokes intensity map at 5 GHz. The rms noise was measured using CASA's IMSTAT task in multiple regions close to the image centre covering several independent beams. We reach 1σ sensitivities of 32 μJy beam⁻¹ on the 1.8 GHz narrow-band image (expected:[3] 32 μJy beam⁻¹), 16 μJy beam⁻¹ on the 1.5-GHz wideband image (expected: 10 μJy beam⁻¹) and 11 μJy beam⁻¹ on the 5-GHz image (expected: 10 μJy beam⁻¹). Image noise and synthesized beam properties are summarized for all three of our observational setups in Table 1, while Fig. 2 shows the final images.

## 3 RESULTS

Previous centimetre continuum observations by Feain et al. (2007) at 6.2 and 8.6 GHz found evidence of a triple radio structure around HE0450−2958. The most westerly component, called 'C3' by Feain et al. (2007), could not be matched with any known extragalactic source (down to an optical limit of ∼26.5 $m_V$ in M05). It is believed to be the synchrotron lobe of an outflow emanating from the quasar, which coincides with the central of the three radio components. The easternmost is associated with the companion SFG and, based on its elongation, is thought to include a contribution from a second, quasar-driven outflow facing in the opposite direction from C3. Our observations extend the ATCA observations by Feain et al. (2007) to longer wavelengths and significantly improve them, both in terms of sensitivity (fourfold) and angular resolution (sixfold) in the *C* band (see Table 1). Due to their higher angular resolution, our observations permit us to map the total extent of the jet and determine the distribution of the synchrotron emission inside the companion SFG and the quasar. With increased sensitivity, we can detect faint substructure in the vicinity of the quasar and companion SFG and identify faint outflow components. Our multiband and higher sensitivity data probe the synchrotron spectrum of the components over a larger range of frequencies, leading to an improved characterization of their spectral shape/index.

Fluxes for the three main emission components were measured interactively with the 2D elliptical Gaussian fitting algorithm included in the CASA Viewer Tool. We also used the CASA task UVMODELFIT with initial estimates based on image plane fits to fit the data with single-component 2D Gaussian model in *uv* space. Both methods yielded highly consistent fluxes and shapes for the quasar, which was later checked by visually inspecting residual images as well. The companion SFG could not be fitted in the *uv* plane due to the relative strength of the surrounding, blended emission, so we used only the central, uncontaminated part of the emission to fit it in the image plane. At 5 GHz (1.8 GHz), we detected the quasar with S/N = 158 (173), the SFG companion with S/N = 93 (86) and the western lobe with S/N = 41 (42). Throughout the paper, we consider sources resolved if their intrinsic source sizes after deconvolution with the restoring beam are not consistent with zero based on source size errors returned by the CASA tasks IMFIT and UVMODELFIT. Using this criterion, both the quasar and the SFG companion were resolved on both *C* band images along both principle axes of the synthesized beam (for derived physical sizes, fluxes and spectral indices, see Table 2).

**Table 2.** HE0450−2958 source properties. B1–B4 are the outflow-related patches of radio emission (see Fig. 4b). Sizes, where given, are the deconvolved, intrinsic, circularized source size FWHMs fitted with CASA. $S_{1.8}$, $S_4$ and $S_6$ were measured on resolution-matched images (see Section 3.1). The spectral index ($\alpha$) was fitted using this resolution-matched photometry. $S_C$ is the flux measured on the combined $C$-band image at 5 GHz with its native, higher resolution. In this image, the flux measured for the quasar and companion SFG is least contaminated by outflow-related emission from B2 to B4; hence, their fluxes are lower than measured at the higher frequency resolution-matched image. Due to its lower S/N in order to achieve the best fit, we measured B1's resolution-matched photometry on the combined, higher sensitivity 5-GHz image. Using less reliable 4 and 6 GHz fluxes for B1, we calculate a consistent but less accurate spectral slope. Fluxes for outflow components B2–B4 were determined on the residual image produced by subtracting the quasar and companion SFG. The spectral index reported for B2–B4 is the average spectral slope of all outflow-related emission features, measured between 4 and 6 GHz (see Section 3.1 for details).

| Object | RA (J2000) | DEC (J2000) | Size (kpc) | $L_{1.4}$ (log(W/Hz)) | $S_{1.8}$ (mJy) | $S_4$ (mJy) | $S_6$ (mJy) | $S_C$ (mJy) | $\alpha$ |
|---|---|---|---|---|---|---|---|---|---|
| Quasar | 04:52:30.10 | −29:53:35.34 | 1 ± 0.1 | 24.23 ± 0.07 | 5.52 ± 0.29 | 2.51 ± 0.15 | 1.77 ± 0.02 | 1.74 ± 0.08 | −1.0 ± 0.1 |
| CSFG | 04:52:30.21 | −29:53:36.27 | 1.3 ± 0.3 | 23.81 ± 0.07 | 2.76 ± 0.27 | 1.39 ± 0.11 | 1.02 ± 0.17 | 0.89 ± 0.02 | −0.8 ± 0.1 |
| B1 | 04:52:30.00 | −29:53:35.26 | – | 23.65 ± 0.14 | 1.34 ± 0.09 | 0.49 ± 0.07 | | 0.45 ± 0.05 | −1.0 ± 0.2 |
| B2 | 04:52:30.05 | −29:53:35.16 | – | 23.14 ± 0.20 | – | – | – | 0.22 ± 0.02 | −0.7 ± 0.3 |
| B3 | 04:52:30.16 | −29:53:35.51 | – | 23.14 ± 0.20 | – | – | – | 0.22 ± 0.03 | −0.7 ± 0.3 |
| B4 | 04:52:30.25 | −29:53:35.41 | – | 22.57 ± 0.22 | – | – | – | 0.059 ± 0.014 | −0.7 ± 0.3 |

Subtracting these high-S/N sources from the 5-GHz[4] image reveals several weaker, but still significant patches of emission around the quasar and the SFG companion. The morphology is not well described by 2D elliptical Gaussians; hence, their fluxes were measured using elliptical apertures and CASA's *imstat* task. Errors for these more irregular emission features were estimated by sampling several emission-free regions with the equally sized apertures and calculating the standard deviation of the fluxes measured in these empty apertures. Their S/N values range from 4 to 10.

## 3.1 Single power-law synchrotron spectra

With our imaging at 1.8, 4 and 6 GHz, we are able to characterize individually the synchrotron spectrum of the quasar and its star-forming companion SFG over a frequency baseline of 4.2 GHz (6.8 GHz including the ATCA *X*-band flux measurements by Feain et al. 2007 at 8.6 GHz). The JVLA images at these three frequencies have a different angular resolution (2.6 arcsec × 0.8 arcsec at 1.8 GHz, 1.2 arcsec × 0.4 arcsec at 4 GHz and 0.7 arcsec × 0.3 arcsec at 6 GHz). For a consistent measurement of the spectral index, we have re-imaged the three frequency data with the restoring beam of the lowest resolution image at 1.8 GHz. We then remeasured integrated fluxes for the three main components of the HE0450−2958 system and calculated the spectral index $\alpha$ defined as

$$S_\nu \propto \nu^\alpha,  \qquad (1)$$

where $S_\nu$ is the flux density at frequency $\nu$.

Fig. 3 shows the synchrotron spectrum of the three main radio components of HE0450−2958 and their fitted spectral slopes. These are highly consistent with the ones derived by Feain et al. (2007, see Table 2); −1.0 ± 0.1 for the quasar, −0.8 ± 0.1 for the companion SFG and −0.9 ± 0.1 for the western lobe (B1). Calculating spectral indices between *L* and *X* bands from Feain et al. (2007), we have −1.0 ± 0.1 for the quasar, −0.8 ± 0.1 for the companion SFG and −1.0 ± 0.1 for the western lobe. The quoted uncertainties are the errors returned by a linear regression fit to the individual flux measurements when these are inversely weighted by their own measurement errors. The observed steep radio spectral index is typical for Seyfert galaxies (e.g. de Bruyn & Wilson 1978; Rush,

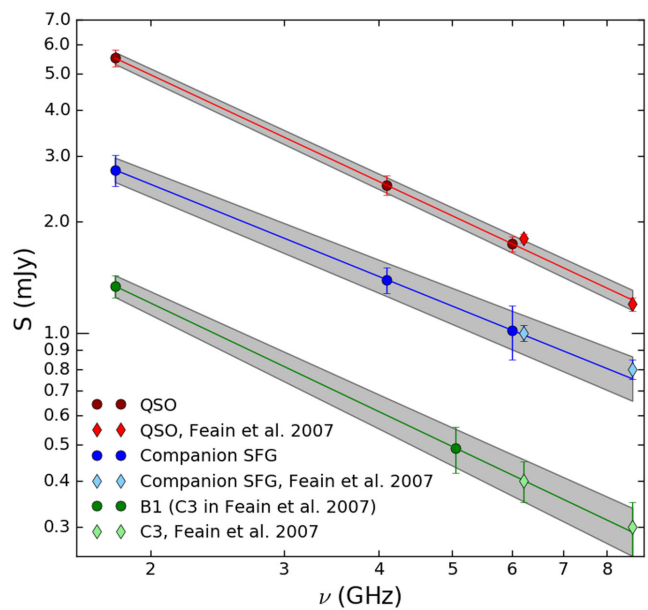

**Figure 3.** Observed radio fluxes of each object as a function of frequency with fitted radio synchrotron spectra. Fluxes were measured on resolution-matched images (see Section 3.1). Dots are fluxes from this work; diamonds are measurements from Feain et al. (2007). Grey areas show 1$\sigma$ confidence intervals. All spectra are consistent with a single power law and match up with the previous flux measurements at >6 GHz from Feain et al. (2007).

Malkan & Edelson 1996). The spectral index of the companion SFG exactly matches the canonical value of an SFG (Condon 1992). The synchrotron spectra of all three main radio sources show no signs of curvature, i.e. spectral indices calculated between 1.8 and 4 GHz, and 4 and 6 GHz are highly consistent, as is the spectral slope between the ATCA 8.6 GHz flux and our lower frequency data. This enables us to reliably extrapolate fluxes measured on the *C*-band image to 1.4 GHz to derive SFRs and to try to constrain AGN contribution to the quasar's radio emission as discussed in Section 4.1.

## 3.2 First evidence for a bipolar outflow in HE0450−2958

The combined, deep 5-GHz intensity map shows substantial residual emission around the two strong single-component sources (see Fig. 2c). To examine these residual features in more detail, we

<hr>

[4] The overall *C*-band flux measurements arising from the combination of the two spectral windows at 4 and 6 GHz.

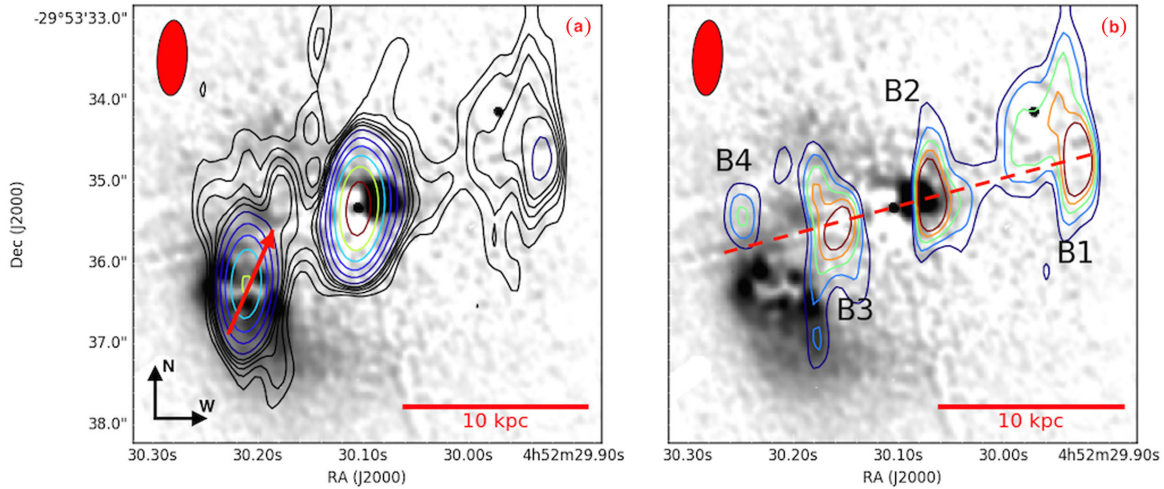

**Figure 4.** *C*-band emission associated with HE0450−2958, superimposed on the PSF-deconvolved, *HST* optical image from M05. Left: JVLA 5 GHz (*C* band) image. Black contours range from 33 μJy beam$^{-1}$ (3σ) to 77 μJy beam$^{-1}$ (7σ) by 1σ steps (same as contours on the right-hand panel for direct comparison). Coloured contours start at 99 μJy beam$^{-1}$, all other levels are $\sqrt{2.5}^n$ multiples of the 9σ contour. To correct for a small astrometric inaccuracy, we have shifted the positions of the radio maps by ∼0.2 arcsec to the south to match quasar's peak position in both images. The centroid of the radio emission from the companion SFG coincides with its dust-obscured central region. Red arrow across the SFG points towards the flatter spectral index area, towards the projected jet axis (dashed line in panel b). Right: residual 5-GHz emission after subtraction of the 2D elliptical Gaussian flux models for the companion SFG and quasar from the overall *C*-band flux distribution shown in Fig. 2(c). Emission features in the residual map (labelled B1 to B4 from west to east) follow a linear alignment reminiscent of a quasar-driven outflow. Contour levels range from 33 μJy beam$^{-1}$ (3σ) to 77 μJy beam$^{-1}$ (7σ) in 1σ steps. Component B2 is spatially associated with an emission line blob detected at optical wavelengths. Component B1 is the western lobe identified by Feain et al. (2007) in lower resolution imaging (see also Fig. 2b and c).

subtracted the fitted 2D elliptical Gaussian flux model of both the quasar and the companion SFG (see introductory paragraphs of Section 3) from the *C*-band image. The residual image contours superimposed on the *HST* optical image from M05 are shown in Fig. 4. To correct for the small astrometric inaccuracy of the optical image, we have shifted the *HST* positions by ∼0.2 arcsec to the south to match quasar's peak position on the JVLA image. The emission line blob found by M05 at optical wavelengths is now detected in radio emission for the first time ('B2'), and is paired with a second radio source ('B3') on the opposite side of the quasar (see Fig. 4). We note that 'B2' was also resolved in the mid-infrared (MIR) by Elbaz et al. (2009); see their fig. 4(a). An imaginary line drawn through these two blobs of emission ends in an extended western radio lobe ('B1'; referred to as 'C3' in Feain et al. 2007) ∼10 kpc to the west of the quasar and intersects the northern quadrant of the companion SFG to the east, as well as a smaller patch of emission,'B4', on the far side of the companion SFG (∼8 kpc east of the quasar). They are most likely physical sources and not imaging artefacts; dirty beam pattern centred on the quasar's and companion SFG's peak positions shows that these residual emission blobs are spatially not correlated with any sidelobes.

This is the first evidence for a bipolar outflow from the HE0450−2958 quasar extending over nearly 10 kpc both westward and in the direction of the companion SFG to the east. Previous observations achieved only a tentative detection of the western lobe but not the eastern component due to blending with the companion SFG's flux in the much more poorly resolved ATCA images.

The 5-GHz radio flux of the outflow components decreases from west to east (i.e. from B1 to B4). B4 on the eastern end of the outflow has a flux density of $59 \pm 14$ μJy (S/N = 4.2), ∼13 per cent of the flux of $450 \pm 46$ μJy (S/N = 9.8) measured for B1 on the opposite side of the outflow. Nearer the quasar, B2 and B3 have very similar flux densities ($220 \pm 24$ μJy for B2 and $220 \pm 32$ μJy for B3; S/N = 9.2 and S/N = 6.9, respectively, and roughly half of B1's flux).

In the individual *C*-band spectral windows, the noise is too high to determine the 4–6-GHz spectral slopes of the outflow components B2–B4 individually. However, it is possible to measure their summed flux at both 4 and 6 GHz with an aperture covering all three objects. The average 4–6-GHz spectral index of the outflow-related emission components B2–B4 is −0.7 ± 0.3.

Using VLT/FORS and VLT/VIMOS data, Letawe et al. (2008) found several clouds of highly ionized gas along the radio axis defined by Feain et al. (2007). With our sub-arcsecond 5-GHz imaging, we can determine a more precise angle for the radio axis, that remains aligned with these clouds. Three pairs of radio emission spatially coincide with these emission line regions. B1, ∼10 kpc from the quasar, is associated spatially with Em3/Em4 and R3 from Letawe et al. (2008). B2, the blob detected 1 kpc to the west of the quasar by *HST*, was found to have AGN-ionized gas according to line diagnostic diagrams. B3 lies too close to the quasar emission, so it was not observed by VLT. B4 (Gal2/Gal3 in Letawe et al. 2008) shows a mix of stars and ionized gas. They have also observed two other emission line regions (Em1 and Em2) ∼10 kpc north and south of the quasar that were not detected by the JVLA at radio frequencies.

## 3.3 Companion galaxy – dust-obscured SF and spectral index variations

Our 5-GHz imaging reveals that the radio emission from the companion SFG peaks in its central, strongly dust-obscured region. This is shown in Fig. 4(a) where radio flux contours are superimposed on the *HST F606W* image of M05 that, at z = 0.285, samples a rest-frame wavelength of 433 nm. The 5-GHz emission associated with the companion SFG is well fit by a single 2D elliptical Gaussian (residuals are <30 μJy) with integrated flux $S_C = 0.89 \pm 0.02$ mJy. The galaxy's apparent source size on the combined 5-GHz image is 0.99 arcsec × 0.44 arcsec. Its emission is resolved on both beam axes with a deconvolved, physical source size (FWHM) of $(1.6 \pm 0.2) \times (1.0 \pm 0.1)$ kpc. After conversion to a 1.4-GHz flux $S_{1.4}$ with equation (1) (where we are able to use our directly measured spectral index $\alpha = -0.8 \pm 0.1$, see Section 3.1), we can estimate the rest-frame 1.4-GHz luminosity of the companion SFG,

$$\left(\frac{L_{1.4}}{\text{W Hz}^{-1}}\right) = 9.51 \times 10^{15} \frac{4\pi}{(1+z)^{(1+\alpha)}} \left(\frac{D_L}{\text{Mpc}}\right)^2 \left(\frac{S_{1.4}}{\text{mJy}}\right),$$
(2)

where $D_L$ is the luminosity distance and $z$ its redshift. Following Bell (2003), the 1.4-GHz luminosity $L_{1.4} = (6.52 \pm 0.20) \times 10^{23}$ W/Hz of the companion SFG translates to an SFR of $360 \pm 11$ $M_\odot$ yr$^{-1}$. This is consistent with the SFR derived from infrared spectral energy distribution (IR SED) fitting in Elbaz et al. (2009) ($\sim$340 $M_\odot$ yr$^{-1}$), who resolved the companion SFG at 11.3 μm using the VLT-VISIR camera (see their fig. 4a). This agreement, in combination with the fact that both radio and MIR emission are resolved on physical scales larger than expected from AGN activity, suggests that the radio synchrotron flux from the companion SFG is dominated by centrally concentrated starburst activity. This makes the presence of a strong AGN in the companion SFG as suggested by Letawe et al. (2009) and Letawe & Magain (2010) unlikely. In Section 4.2.1, we attempt to find an upper limit to a possibly weaker AGN activity.

In Section 3.2, we showed that outflow-related radio emission features are linearly aligned along an axis that crosses the northern part of the companion SFG. We lack information on the 3D orientation of the outflow to unambiguously tell whether it actually interacts with the companion SFG. However, the marked flux difference between the two most distant outflow components, B1 and B4 (which is in contrast with the very similar brightness of components B2 and B3, that are located closer to the quasar), indicates that the density of the medium eastern and western lobes traverse through is not the same. This could be interpreted as evidence that the eastern outflow does impinge on the companion SFG and interact with its ISM. If this is the case, we might expect a different synchrotron spectral index in the northern sector of the companion SFG due to the acceleration of charged particles in shocks in the jet–ISM interaction region. We will discuss the implications of this for the jet-induced SF hypothesis in Section 4.2.

Given that we have spatially resolved the companion SFG in our *C*-band image, we made a first attempt to search for internal, coherent spectral index variations across the galaxy. We consider all pixels where flux is detected at S/N > 3 in both the spectral window centred at 4 and at 6 GHz, an area equivalent to 3–4 independent resolution elements (referenced to the average synthesized beam size of the two spectral windows). It is possible to define up to three independent, beam-sized regions on the image to sample spectral slope variations between the resolution-matched 4- and 6-GHz images (common resolution: 1.16 arcsec × 0.41 arcsec). The first of these beam-sized regions was centred on the peak of the 2D Gaussian model for the companion SFG. The other two regions were offset 0.8 arcsec (3.5 kpc) from the galaxy centre, along an axis running approximately north–south across the galaxy. (This axis links the regions nearest and farthest in projection from the quasar outflow axis and is indicated with a red arrow in Fig. 4a.) The head (tail) of the arrow coincides with the centre of the southern (northern) resolution element used in our spectral index analysis.) The northernmost resolution element may contain some emission from the quasar outflow component B3. We thus also used a two-region configuration (where independently resolved regions were centred on the southern and the northern half of the galaxy along the same axis as before, see red arrow on Fig. 4a) as a second, more conservative test of spectral index variations out to maximally 1.8 kpc from the centre. The average spectral indices measured pixel-to-pixel in the three-beam configuration are $-1.35^{+0.81}_{-0.38}$, $-0.93^{+0.10}_{-0.11}$ and $-0.49^{+0.58}_{-0.95}$, in the two-beam configuration $-1.07^{+0.37}_{-0.14}$ and $-0.79^{+0.15}_{-0.26}$ (measurements always ordered from south to north, see in Fig. 4a the red arrow). These uncertainties span the 1$\sigma$ confidence region and were estimated using a resampling approach to generate a million realizations of the individual pixel fluxes in each spectral window based on the noise measured at 4 and 6 GHz. While these values suggest a consistent trend towards a flatter synchrotron spectrum in northern part of the companion SFG, the associated uncertainties imply that the spectral index in the northern- and southern-most resolution elements is still consistent within their errors. To test at which significance level we are able to constrain the spatial spectral index gradient from north to south, we derived the best-fitting parameters of the following linear relation between spectral index $\alpha(r)$ and the distance $r - r_c$ from the central galaxy position (determined by the distances to the central position of each of the regions used):

$$\alpha(r) = \alpha(r_c) + \text{d}\alpha/\text{d}r \times (r - r_c).$$
(3)

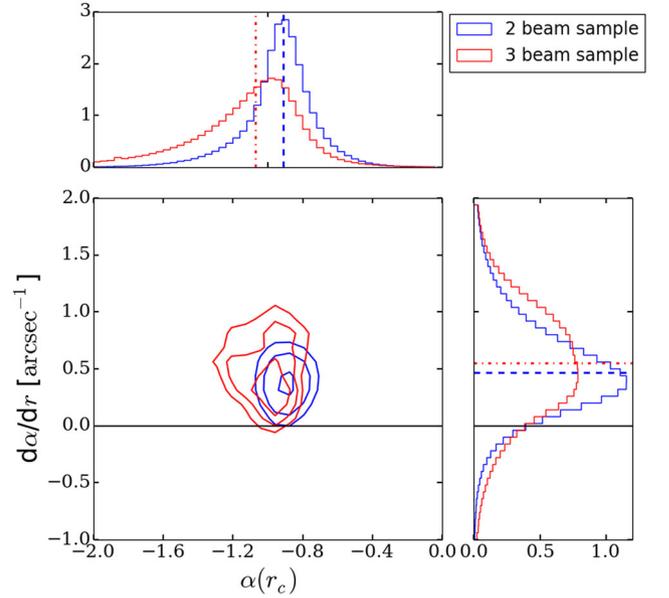

**Figure 5.** 2D histogram of measured central spectral indices ($\alpha(r_c)$) and spectral index gradients (d$\alpha$/d$r$) across the star-forming companion SFG using two (blue) and three (red) beam-sized regions to sample fluxes and measure spectral indices (for details see Section 3.3). Contours are at 50 per cent, 68 per cent and 98 per cent of the maxima of each distribution. Dotted lines show the medians of the distributions. Both spatial sampling approaches yield consistent results; the two beam sampling has more variance. 91 per cent (two-beam configuration) and 85 per cent (three-beam configuration) of the spectral slope gradients are consistent with a flatter spectral index in the northern part of the galaxy (d$\alpha$/d$r$ > 0) that overlaps with the (projected) outflow axis.

Here $\alpha_{(r=r_c)}$, the spectral index at $r_c$, and d$\alpha$/d$r$, the spatial spectral index gradient, are the two free parameters. The results are shown in Fig. 5. The best-fitting central spectral index values $\alpha_{(r=r_c)}$ are $-0.9^{+0.0}_{-0.2}$ (two beam sampling) and $-1.1^{+0.2}_{-0.4}$ (three beam sampling). Offsets from the measured $-0.8 \pm 0.1$ (reported in Section 3.1) are the consequence of the different resolutions used to extract the fluxes, and that these spectral indices were measured over a shorter frequency baseline (1.8-GHz data could not be used in the present analysis due to its poorer resolution). We find a median value of $d\alpha/dr = 0.11^{+0.10}_{-0.08}$ kpc$^{-1}$ (two-beam configuration) and $0.13^{+0.12}_{-0.11}$ kpc$^{-1}$ (three-beam configuration) for the gradient of the spatial spectral index measurements. (The $1\sigma$ errors quoted are the result of selecting one of our resampled spectral index values in each of the three/two resolution elements and fitting equation 3 for all our Monte Carlo realizations.) 91 per cent (two-beam configuration) and 85 per cent (three-beam configuration) of the values lie above 0, i.e. imply a flattening of the spectral index in the companion SFG's northern sector. With the present data, we have only been able to detect these spatial spectral index variations within the companion SFG at a low-significance level. Taken at face value, however, the flatter spectral index in the northern part of the galaxy could be interpreted as a younger synchrotron-emitting electron population in this region. This observation is compatible with the jet-induced SF hypothesis of Elbaz et al. (2009, see further discussion in Section 4.2) and may hence be taken as tentative evidence that the outflow identified in Section 3.2 not only overlaps with the companion SFG in projection, but is actually interacting with its ISM.

### 3.4 Quasar – evidence for a star-forming host galaxy and classification as a compact steep spectrum source

The measured flux of the quasar at the native 5-GHz resolution is $1.74 \pm 0.08$ mJy, corresponding to $\log(L_{1.4}[\text{W/Hz}]) = 24.23 \pm 0.07$ at the quasar redshift. The deconvolved, physical source size of the radio emission associated with the quasar derived from the 5-GHz image is $(1.2 \pm 0.3) \times (0.8 \pm 0.3)$ kpc. This is consistent with the 600 pc lower limit of Feain et al. (2007). The extent of the emission suggests that not all of it is due to the central engine; it has at least some contribution from SF activity as well. The large-scale environment (>20 kpc) shows no sign of extended emission line regions (see Fig. 2a) down to the $3\sigma$ level of 48 mJy beam$^{-1}$.

Based on its steep radio spectrum (see Section 3) and compact linear projected size of its outflow ($\sim$20 kpc across, see Section 3.2), the quasar can be classified as a compact steep spectrum source (O'Dea 1998) that may imply that it is a recently activated, nascent AGN (see e.g. Fanti et al. 1995). The absence of far-field ionized regions supports this interpretation as well.

## 4 DISCUSSION

### 4.1 Disentangling SF and AGN activity in the quasar

The physical extent of the emission towards the quasar suggests that the putative quasar host contains at least some SF activity (see Section 3.4). To gauge its relative importance compared to the AGN component, we use archival VLBI data. We will follow a similar line of argument as Feain et al. (2007) but can now base our statements on the more accurately determined radio spectral indices of the quasar and surrounding emission line regions (see Section 3.1). HE0450−2958 was observed in 1990 and 1991 (Roy et al. 1994) with the decommissioned Parkes–Tidbinbilla real-time interferometer (PTI; Norris et al. 1988) at 2.3 GHz. At this frequency, the interferometer was only sensitive to emission from scales smaller than 0.1 arcsec ($\sim$400 pc at the redshift of HE0450−2958), e.g. compact emission from AGN. HE0450−2958 was not detected down to a $3\sigma$ limit of 2 mJy.

This upper limit can be used as a constraint on the radio flux from the central AGN, and hence to calculate a lower limit on the spatially more extended SF activity of the host galaxy. Our best estimate of the combined flux due to AGN and SF activity in the quasar comes from our sub-arcsecond *C*-band image. To derive a constraint on the host galaxy SFR, we thus have to extrapolate from the PTI 2.3 GHz upper detection limit to a flux constraint at 5 GHz. While we do not have accurate information on the spectral index of the AGN-related emission, the overall spectral shape of the quasar, as measured in the resolution-matched 1.8-, 4- and 6-GHz images (see Section 3.1), provides some clues. The total quasar flux on these images at each frequency has contributions from AGN-related emission, SF activity and the quasar-driven outflow components B2 and B3 with unknown proportions, but none of them are negligible. As shown in Section 3.1, components B2 and B3 display a similarly steep spectral index $\alpha_{B2/B3} = -0.7 \pm 0.3$, which is similar to that expected for the host galaxy SF contribution. Consequently, the AGN-related emission most likely has a spectrum comparable to all other components in order for the simple power-law shape of the overall quasar spectrum ($\alpha_{\text{quasar}} - 1 \pm 0.1$, see Table 2) to be preserved. In this case, extrapolation of the PTI non-detection at 2.3–5 GHz and subsequent subtraction from the 2D Gaussian flux distribution centred on the quasar formally suggests that the quasar host galaxy could be undergoing SF activity approaching the level of a ULIRG (SFR $\geq 100 \, M_\odot \, \text{yr}^{-1}$) and hence similar in intensity to the companion SFG.

It is worth noting that, according to Papadopoulos et al. (2008), the quasar host displays approximately fivefold weaker molecular emission than the companion SFG; for a common velocity range and CO-to-H$_2$ conversion factor ($\alpha_{CO} \approx 0.8 \, M_\odot$ [K km s$^{-1}$ pc$^2$]$^{-1}$), they estimated a $3\sigma$ upper limit on the molecular gas mass of $3.6 \times 10^9 \, M_\odot$. The halo of ionized gas surrounding the quasar (Letawe et al. 2008) is evidence of significant energy input from the AGN, in the form of radiative feedback, that may also have contributed to depleting the gas reservoir of the quasar host galaxy. At least two parallel lines of evidence (signatures of strong AGN feedback, plus matching SFR estimates for the companion SFG from spatially resolved radio data and unresolved FIR, see Section 4.2.1) hence suggest that ULIRG-like activity should be regarded as an upper limit to the host galaxy SFR. We also note that the PTI constraint from a single baseline snapshot observation might be sub-optimal due to its poor uv coverage.

In conclusion, due to our inability to robustly separate AGN and SF activity in the quasar host galaxy, its properties remain poorly known. To directly determine these, new high-resolution observations are needed. A more sensitive VLBI follow-up could determine the amount of nuclear emission in the quasar. With ALMA, one could obtain more sensitive CO observations to place a more stringent limit on the gas content and a measurement of the dust continuum to see if there is evidence for cool dust heated only by SF in the quasar host. Moreover, high-resolution ALMA imaging also has the potential to clarify whether the blob B2 contains dust-obscured SF activity as traced by our radio detection and by MIR in Elbaz et al. (2009) or if it represents AGN-heated dust by an outflow component we have detected at radio wavelengths, as suggested by Letawe et al. (2008).

### 4.2 The companion galaxy – type and origin of activity

In previous work on HE0450−2958, two distinct triggering mechanisms for the strong SF activity in the companion SFG have been proposed: (a) interaction/merging between the companion and the quasar host galaxy (Letawe et al. 2008; Papadopoulos et al. 2008; Letawe & Magain 2010), and (b) jet-induced SF (Feain et al. 2007; Elbaz et al. 2009). We will now discuss whether our new JVLA observations provide evidence in favour of either of these scenarios. In the preceding section, we showed that the quasar host, in addition to the AGN activity, is most likely undergoing SF as well. Our spatially resolved JVLA imaging allows us to address whether AGN and SF activity might be occurring simultaneously also in the companion SFG.

### 4.2.1 The relative importance of SF and AGN activity in the companion SFG

Based on its 5 GHz flux, we infer an SFR of $360 \pm 11 \, M_\odot \, yr^{-1}$ for the companion SFG. This agrees well with the previously reported values of $\sim 370 \, M_\odot \, yr^{-1}$ (Papadopoulos et al. 2008) and $\sim 340 \, M_\odot \, yr^{-1}$ (Elbaz et al. 2009), which were derived with SED fits predominantly based on MIR and FIR flux densities measured with *IRAS* and VLT-VISIR. Since the companion SFG resides squarely on the IR–radio relation ($q = 2.5$, i.e. offset by a mere 0.02 dex from the local IR–radio relation in Yun, Reddy & Condon 2001 and Bell 2003), there is no clear evidence of AGN activity contributing to the radio or IR flux densities of the companion SFG. However, given our detection of spatially extended SF in the quasar host galaxy (see Section 4.1), it is plausible that it could potentially contribute to the *IRAS* FIR photometry at levels comparable to the contribution from the companion SFG. In the simultaneous, joint IR SED fitting of the companion SFG and the quasar, the companion galaxy was previously assumed by both Papadopoulos et al. (2008) and Elbaz et al. (2009) to strongly dominate the *IRAS* photometry[5] at rest-frame wavelengths >40 μm by both Papadopoulos et al. (2008) and Elbaz et al. (2009). With a more even split of dust-obscured SF activity between quasar host and companion SFG (see Section 4.1), the SFR values reported for the companion SFG by Papadopoulos et al. (2008) and Elbaz et al. (2009) would effectively represent upper limits, implying that its radio-based SFR exceeds that inferred from the FIR. While this could be interpreted as due to contamination from AGN-related radio emission, it is not clear that the effect would be large enough to move the companion SFG significantly beyond the scatter of the IR–radio correlation (∼0.3 dex; e.g. Yun et al. 2001; Sargent et al. 2010) such that the companion SFG would enter the regime of radio-loud AGN. Furthermore, we have shown in Section 3.3 that the radio continuum emission from the companion SFG is distributed on a scale of ∼1.5 kpc. We note that on the 11.3-μm VLT-VISIR image, the companion SFG was also found to be extended (i.e. there is no evidence for a dominant near-IR power-law AGN component; Elbaz et al. 2009), which further implies that the emission is mostly due to SF activity in companion SFG. For the remainder of the discussion, we will hence assume that the SFR of the companion SFG is not changed significantly by assigning some FIR flux to the quasar host galaxy and adopt $360 \, M_\odot \, yr^{-1}$ for its SFR.

This estimate on the SFR allows us to reassess the companion SFG's specific SFR and depletion time-scale.[6] The stellar mass of the companion SFG is estimated at $5–6 \times 10^{10} \, M_\odot$ by Elbaz et al. (2009). Its specific SFR ($\sim 6.5 \, Gyr^{-1}$) makes it a strong starburst galaxy with a factor of 40 offset from the main sequence (MS; using the evolutionary fit to compiled literature data) of SFG (Sargent et al. 2014) at $z \sim 0.3$. Elbaz et al. (2009) also found that it has a young (40–200 Myr) stellar population.

The gas mass of the companion SFG was estimated[7] by Papadopoulos et al. (2008) to be $1.3–2.3 \times 10^{10} \, M_\odot$. In combination with the radio-based SFR, these gas masses translate to a gas depletion time-scale of $t_{dep} \equiv M_{H_2}/SFR = 12–50 \, Myr$, assuming a constant gas consumption rate and no in- or outflows. Together with the young stellar population and the large offset from the star-forming MS, these short depletion times indicate that the companion SFG is currently undergoing a short-lived, high-efficiency SF phase, which might provide further clues towards the nature of the triggering mechanisms, as discussed in the following section.

### 4.2.2 Starburst in the companion SFG – merger- or jet-induced origins?

In summary, the observations we have are: (i) projected outflow axis intersecting companion SFG's northern half (Section 3.2), (ii) possible spectral index variation across the star-forming region in the companion SFG towards the point of intersection (Section 3.3), (iii) a short, very intense starburst phase in companion SFG (Section 4.2.1) and (iv) a small projected linear size for the outflow from the quasar, indicative of recently started AGN activity (Section 3.2).

*Merger-induced SF*    Should the evidence for (ii) not be corroborated by future, higher quality observations and (i) be purely due to projection or if the jet–ISM interaction does not trigger SF in the companion SFG, then an interaction-induced starburst is the most plausible mechanism for generating SF with the intensity observed in this galaxy. Specifically, Papadopoulos et al. (2008) referred to this process in HE0450−2958 as a wet–dry merger due to the strongly asymmetric distribution of gas between the quasar host (which remained undetected in their CO follow-up observations) and its neighbouring galaxy. Merger-driven accumulation and compression of the dissipative gas phase in the central region of merger remnants is a well-established and widespread triggering mechanism in low-redshift, dust-obscured starbursts (Sanders & Mirabel 1996, and references therein). However, in local (U)LIRGs, starburst activity is predominantly observed in the nuclei of fully coalesced galaxies (Sanders et al. 2003), while in the case of HE0450−2958, the ULIRG phase already occurs when the interacting galaxies are still separated by ∼7 kpc. We also note that wet–dry interactions of $0 < z < 1.2$ galaxies on average do not enhance SF activity (Hwang et al. 2011). A possible explanation for this is the removal of gas from the late-type galaxy involved in wet–dry mergers through stripping in the hot halo of the early-type merging partner. Letawe et al. (2008) suggest that such a hot, ionized gas halo also surrounds HE 0450−2958, probably due to radiative feedback from the AGN. There are hence at least two atypical features of HE 0450−2958 if a wet–dry merger is indeed the main cause for the SF activity in the companion SFG.

---

[5] Due to the relatively poor angular resolution of *IRAS*, the emission from the quasar and companion SFG is entirely blended in the *IRAS* maps.

[6] We note that even for a ULIRG-like SFR for the quasar host (SFR $\gtrsim 100 \, M_\odot \, yr^{-1}$), which would lead to a commensurate reduction of SFR of the companion SFG, our subsequent conclusions on depletion time and excess specific SFR remain qualitatively correct. (That is, the statement that the companion SFG lies well above the star-forming MS and has a short depletion time compared to MS galaxies is robust to such changes.)

[7] Papadopoulos et al. (2008) adopt a CO-to-$H_2$ conversion factor in the range $\alpha_{CO} = 0.55–1 \, [M_\odot \, K \, km \, s^{-1} \, pc^2]^{-1}$, where the lower limit is for optically thin CO[1–0] emission and the upper bound reflects dynamical constraints derived by Downes & Solomon (1998) for local starburst ULIRGs. Following the scaling relations calibrated in the 2-Star Formation Mode framework of Sargent et al. (2014), a similar value of $\alpha_{CO} = 0.8–1 \, [M_\odot \, K \, km \, s^{-1} \, pc^2]^{-1}$ is on average expected on statistical grounds for a $z \sim 0.3$ galaxy, with the stellar mass and SFR measured for the companion SFG. Note that larger $\alpha_{CO}$ values we found in two-phase modelling of local starbursts (Papadopoulos et al. 2012) correspond to depletion times of $\sim 100 \, Myr$ for the companion SFG. This figure remains significantly shorter than the depletion times of normal low-$z$ galaxies (Leroy et al. 2008; Saintonge et al. 2011). Since both the quasar and the companion SFG are ULIRGs, this alternative technique for deriving CO-to-$H_2$ conversion factors is unlikely to significantly change the relative amount of molecular gas in the two systems.

*Jet-induced SF hypothesis* Evidence for jet-induced SF has so far been reported for a fairly small number of galaxies: Minkowski's Object, a peculiar star-forming dwarf galaxy lying in the cone of the jet from the powerful radio source NGC 541 (Croft et al. 2006); Centaurus A (Schiminovich et al. 1994; Charmandaris, Combes & van der Hulst 2000) and 3C 285 (van Breugel & Dey 1993), both nearby radio galaxies with complex, star-forming filamentary structures along their radio lobes; and 4C 41.17, a radio source at $z = 3.8$ (Bicknell et al. 2000; de Breuck et al. 2005; Papadopoulos et al. 2005). Their rarity and the low redshift of most of these objects reflects the fact that high-resolution and detailed multifrequency data are required to identify such systems. However, jet-induced SF may be more common, especially among high-redshift galaxies (as suggested by, e.g. radio-optical alignments, see McCarthy et al. 1987).

In the case of HE 0450−2958, the analysis of the spatial distribution of emission line ratios in Letawe et al. (2008) revealed the presence of shocked gas in the northern half of the companion SFG consistent with a jet–ISM interaction in the companion SFG as evidenced by (i) and (ii). It is possible for such an interaction to accelerate the conversion of H I to H$_2$ as discussed by, e.g. Nesvadba et al. (2011). For this H$_2$ reservoir to be able to actually produce new stars, it needs to cool and reach sufficiently high densities. Cooling may only be possible where turbulent energy input by the jet does not dominate the ISM energetics (e.g. after the jet has subsided or at a sufficient distance from the interaction region). For high-efficiency starburst activity as observed in the companion SFG, a significant fraction of the gas must, furthermore, be in a dense phase (Renaud, Kraljic & Bournaud 2012). It is possible that these conditions are met in its central region, i.e. >1 kpc from the jet axis and where the gas density would have been highest to begin with. Alternatively, the central parts of the companion SFG might have been impacted more directly by the jet in the past due to either the relative motion of the companion with respect to the quasar (which, for purely tangential motion of the order of 100 km s$^{-1}$, could have caused the jet axis to wander from the galaxy centre to its present location on a time-scale of ∼$10^7$ yr) or quasar jet wobbling.

With the currently available data, it is not possible to draw a definitive conclusion on whether an interaction-induced starburst, or SF related to a jet–ISM interaction, is the primary trigger of the activity in the companion SFG. It is even conceivable that a combination of these two phenomena might be occurring in HE 0450−2958. It should be mentioned, however, that the probability of two extremely short-lived and rare events such as (iii) and (iv) happening simultaneously in the same system without any causal connection seems small. Intriguingly, the SF efficiency we have re-estimated based on our new JLVA continuum data is also comparable to that found for Minkowski's Object (depletion time <20 Myr; Salomé, Salomé & Combes 2015).

The continued ambiguity of the SF process in the companion SFG illustrates how complex (combinations of) feedback processes are in practice and how important in-depth observations are for their implementation in galaxy evolution models. For an improved understanding of this gas-rich and dusty system, further study of the gas kinematics and energetics is indispensable. Specifically, a spatially resolved imaging of the CO spectral line excitation diagram would permit a separate mapping of the excitation, density, and kinematics of both the warm and cool molecular reservoirs. Moreover, follow-up of a dense gas tracer such as hydrogen cyanide (HCN) could reveal whether or not the HCN/CO ratio of the companion SFG is comparable to that of local, merger-driven starburst ULIRGs.

# 5 SUMMARY

We have observed HE 0450−2958, a $z = 0.285$ galaxy pair consisting of a quasar with an elusive host galaxy and an actively star-forming companion SFG, at 1.8, 4 and 6 GHz using the JVLA.

We find that radio emission towards the quasar is resolved. The intrinsic physical size of the radio-emitting region is ∼1 kpc. The presence of this extended emission suggests that the AGN host galaxy is still forming stars and might be doing so at a rate of $\gtrsim 100 \, M_\odot \, yr^{-1}$.

Our 5-GHz *C*-band radio image (angular resolution 0.9 arcsec × 0.4 arcsec) provides the first evidence of a bipolar outflow (linear projected extent ∼20 kpc) that is aligned with the northern quadrant of the companion SFG. The companion SFG is resolved at 4 and 6 GHz, enabling us to map variations of the synchrotron spectral index within the companion SFG and, in particular, towards the point of intersection between the jet axis and the galaxy. We find tentative evidence for a flattening of the spectral index in the potential interaction region. Our highest resolution image shows that SF activity (with spatial extent ∼1.4 kpc) is located in the dust obscured, central part of the galaxy. We find no evidence for significant AGN activity in the companion SFG, which is undergoing a short (gas depletion time-scale of the order of 10 Myr) and very intense (SFR ∼360 $M_\odot \, yr^{-1}$) starburst phase.

In previous literature, both a galaxy–galaxy interaction and a jet-induced SF have been proposed as triggering mechanisms for the starburst in the companion galaxy. We revisit the evidence for both of these in the light of our new JVLA observations in Section 4.2. These observations were designed to test the jet-induced SF hypothesis and the spatial alignment of the quasar jet axis and the companion galaxy, combined with the tentative evidence for radio spectral index variations within it, implies that a jet–ISM interaction remains a valid scenario. H I to H$_2$ conversion via turbulent compression could then have led to SF in regions where energy input from the jet has subsided enough to allow cooling. With the currently available data, it is not possible to exclude interaction-driven gas dynamics as the prime trigger of the starburst. It is, however, one of the best-known examples of strong SF activity in the presence of a quasar-driven jet. Our new observations make HE 0450−2958 a strong candidate for jet-induced SF and one of the rare links between local systems (like Minkowski's Object or Centaurus A) and the high-$z$ regime where radio-optical alignments suggest that this phenomenon could be more common. HE 0450−2958 is also an excellent illustration of the complex interplay between different astrophysical processes: simultaneous AGN and SF activity within individual galaxies and a combination of both negative (quasar host) and positive feedback (companion SFG). Further study of systems like this will provide important clues for understanding all facets of AGN feedback processes.

## ACKNOWLEDGEMENTS

We thank the anonymous referee for a helpful report that allowed us to improve the manuscript. We thank Arwa Dabbech, Chiara Ferrari, Ivan Marti-Vidal, Stephen Wilkins and Ilana Feain for useful discussions. DM acknowledges support from the Science and Technology Facilities Council (grant number ST/M503836/1). MTS acknowledges support from a Royal Society Leverhulme Trust Senior Research Fellowship. PPP is supported by an Ernest Rutherford Fellowship. JS was supported in part by ERC Project No. 267117 (DARK) hosted by Université Pierre et Marie Curie (UPMC) – Paris 6. The National Radio Astronomy Observatory is a facility of the National Science Foundation operated under cooperative agreement by Associated Universities, Inc. This research made use of APLPY, an open-source plotting package for PYTHON hosted at http://aplpy.github.com. This research made use of AS-TROPY, a community-developed core PYTHON package for Astronomy (Astropy Collaboration 2013).

This paper has been typeset from a TeX/LaTeX file prepared by the author.